\documentclass[11pt]{article}
\usepackage{moriond,epsfig}

\def\Journal#1#2#3#4{{#1} {\bf #2}, #3 (#4)}

\def\NCA{\em Nuovo Cimento}
\def\JHEP{\em JHEP}

\def\NPB{{\em Nucl. Phys.} B}
\def\PLB{{\em Phys. Lett.}  B}
\def\PRL{\em Phys. Rev. Lett.}
\def\PRD{{\em Phys. Rev.} D}

\def\beq{\begin{equation}}
\def\eeq{\end{equation}}
\def\bea{\begin{eqnarray}}
\def\eea{\end{eqnarray}}

\def\bmeg{\mathrm{BR}(\mu\to e\gamma)}
\def\simlt{\stackrel{<}{{}_\sim}}
\def\simgt{\stackrel{>}{{}_\sim}}   
\def\gev{\,\mathrm{GeV}}
\def\teg{\tau\to e\gamma}
\def\tmg{\tau\to\mu\gamma}

\newcommand{\ynu}[1]{\mathbf{Y}_{\nu}^{#1}}
\newcommand{\mm}[1]{\mathbf{M}_{#1}}
\newcommand{\OO}[1]{\mathbf{\Omega}_{#1}} 
\newcommand{\Un}[1]{\mathbf{U}_\nu^{#1}} 
\begin{document}
\vspace*{4cm}
\title{DECOUPLING, LEPTON FLAVOUR VIOLATION AND LEPTOGENESIS}

\author{K.\ TURZY\'NSKI}

\address{Institute of Theoretical Physics, Warsaw University,\\
ul.\ Ho\.za 69, PL-00-681 Warsaw, Poland}

\maketitle\abstracts{
Several classes of neutrino seesaw mass models, which can naturally account for hierarchical neutrino masses and the bi-large pattern of neutrino mixing, are constructed from a bottom-up perspective based on the idea of decoupling of one right-chiral neutrino from the seesaw mechanism. The interplay between the predictions for the lepton-flavour violation in the MSSM with universal soft masses and for leptogenesis is studied. In particular, classes of neutrino mass models in which successful low-temperature leptogenesis implies $\mathrm{BR}(\mu\to e\gamma)$ potentially observable in upcoming experiments are identified.}

\section{Introduction}

The discovery of neutrino flavour transitions,
whose only explanation are neutrino oscillations,
necessarily implies that neutrinos have masses, much smaller than the masses of other
elementary particles, and a bi-large pattern of mixing, very different
from that of quarks. These results cannot be accommodated in the Standard
Model of particle physics, in which, by construction, the neutrinos
are massless and each of the partial lepton numbers $L_e$, $L_\mu$ and 
$L_\tau$ is separately conserved. Hence, neutrino experiments provide
unquestionable empirical evidence for the necessity of extending
the Standard Model. This necessity is supported by a serious theoretical
argument that the explanation of the huge hierarchy between the
energy scale of weak interactions and the Planck scale 
(or, in general, a high scale of new physics)
requires
an enormous fine-tuning between the tree-level parameters of the
Standard Model and quantum corrections. This hierarchy problem
can be solved by introducing supersymmetry, softly broken by masses
of new particles of order of $1\,\mathrm{TeV}$.

The smallness of neutrino masses can, in turn, be elegantly explained
by the seesaw mechanism\cite{seesaw}, 
postulating that they are generated in
Yukawa interactions, involving an exchange of very heavy Majorana fermions,
the right-chiral neutrinos. 
Their interactions can be described by the following Lagrangian:
\beq
\label{potential}
\Delta\mathcal{L} = - \epsilon_{ij}H_i N_{K} \ynu{KA} \ell_{Aj} - \frac{1}{2} \mm{KL} N_{K} N_{L} + \mathrm{H.c.}~,
\eeq
where $H$ and $\ell_{A}$ are Higgs and lepton doublets, 
$N_{K}$, ${K}=1,2,3$ 
denote the right-chiral neutrinos, 
$\mm{KL}$ is their Majorana mass matrix 
and $\ynu{KA}$ is the matrix of the neutrino Yukawa couplings. 
Integrating out the heavy fields $N_K$ provides 
(after the electroweak symmetry breaking) 
the remaining neutrinos with small Majorana masses:
\beq
\label{seesawfor}
\mathbf{m}_\nu = -\langle H\rangle^2\mathbf{Y}_\nu^T \mathbf{M}^{-1} \ynu{}~,
\eeq
where $\mathbf{m}_\nu$ has eigenvalues $m_{\nu_1}$, $m_{\nu_2}$ and $m_{\nu_3}$, and $\langle H\rangle$ denotes the vacuum
expectation value of the relevant Higgs field.

The seesaw mechanism introduces new parameters
to the Standard Model (or to its supersymmetric extension, the MSSM),
which cannot be fully determined even with infinitely accurate measurements
of neutrino masses, mixing angles and $CP$ properties. One can attempt
to reduce the number of unconstrained parameters
by building appropriate neutrino mass models,
e.g.~by postulating that the hierarchical neutrino masses and the bi-large
pattern of neutrino mixing is generated without fine-tuning in the seesaw
mechanism. This can be achieved by assuming that one right-chiral
neutrino gives very small contributions to the mass matrix of the light
neutrinos, i.e.~it decouples from the seesaw mechanism\cite{kingall}. 
In our analysis, we have used the Casas-Ibarra parametrization of the neutrino
Yukawa matrix\cite{casas01}, which directly links the neutrino observables to the parameters
of the seesaw mechanism, thereby allowing to use the observables as
input parameters for building neutrino mass models
\beq
\ynu{AB} = i \frac{M_A^{1/2}}{\langle H\rangle}\sum_D \OO{AD}m_{\nu_D}^{1/2} {\Un{BD}}^\ast \qquad {\OO{}}^T\OO{}=\mathbf{1}~,
\label{YukawaOmega}
\eeq
where $M_A$ is the mass of $N_A$ and $\Un{}$ is the neutrino mixing
matrix.
This parametrization takes a particularly simple form 
in neutrino seesaw mass models with one right-chiral neutrino
decoupled:
\beq
\OO{} \approx \left( \begin{array}{ccc} 1 & 0 & 0 \\ 0 & z & p \\ 0 & \mp p & \pm z \end{array}\right), \left( \begin{array}{ccc}  0 & z & p \\ 1 & 0 & 0 \\ 0 & \mp p & \pm z \end{array}\right) \textrm{ and }\left( \begin{array}{ccc}  0 & z & p \\ 0 & \mp p & \pm z \\ 1 & 0 & 0 \end{array}\right)~,
\eeq
for $N_1$, $N_2$ and $N_3$ decoupled, respectively. 

Due to the Majorana nature of the right-chiral neutrinos,
the total lepton number $L=L_e+L_\mu+L_\tau$ is violated in their
interactions. At presently accessible energies, the only experimental
sign of $L$ violation would be the observation of neutrinoless double
beta decay\cite{dbeta},
since the right-chiral neutrinos are probably too heavy to be produced
in any terrestial experiments. They could have been, however, fairly
abundant in the early Universe, if it was sufficiently hot, and their
$CP$ violating decays could have produced a lepton asymmetry, subsequently
transformed, thanks to sphaleron transitions, into the baryon asymmetry
observed today (baryogenesis via leptogenesis)\cite{bvl}. 

In the MSSM embedded in supergravity, 
there is, however, an upper bound on the temperature of
the Universe after inflation, the so-called reheating temperature 
$T_\mathrm{RH}$. If $T_\mathrm{RH}$ is too high, too many gravitinos are produced
and they make up too much dark matter or their late decays destroy the
observationally confirmed predictions of the primordial nucleosynthesis.
Depending on the mass spectra of supersymmetric particles, the bound
on the reheating temperature can be strong\cite{kohri01} ($T_\mathrm{RH}\simlt 10^{7-8}\gev$)
or weak\cite{roszkowski04} ($T_\mathrm{RH}\simlt 10^9\gev$).

It is interesting to study a possible interplay
between
lepton-flavour violating (LFV) decays of the charged leptons in the MSSM
and 
the violation of the total lepton number in
leptogenesis.
As we shall show, in some neutrino mass models,
the requirement of successful
leptogenesis 
consistent with the gravitino bounds
on the reheating temperature
can give some predictions 
for the rates of the LFV decays of charged leptons
and reduce the ambiguity in determining the neutrino
couplings from the low-energy experiments.

\section{LFV radiative decays of the charged leptons}

In the MSSM extended with the seesaw mechanism, even if the mechanism
of supersymmetry breaking does not introduce by itself any flavour violation
(e.g.~as in mSUGRA with sfermion mass matrices 
$\tilde{\mathbf{m}}_f^2=m_0^2\mathbf{1}$, gaugino masses $M_{1/2}$ 
and the three-scalar couplings proportional to the relevant Yukawa
matrices, $\mathbf{A}_f=A_0\mathbf{Y}_f$, at the scale of Grand
Unification),
there are flavour-violating quantum corrections to the mass matrix of
sleptons, the scalar partners of leptons.
These corrections\footnote{For some values of $m_0,M_{1/2}$,
the approximation in Eq.~\ref{RGEsol1} does not work well and
corrections beyond the leading logarithm have to be 
included\cite{chankowski04,petcov04}.}, 
\begin{equation}
\left(\tilde{m}_L^2\right)_{AB}\approx -\frac{3m_0^2+A_0^2}{8\pi^2} \sum_{K=1}^3 \mathbf{Y}_\nu^{KA\ast} \mathbf{Y}_\nu^{KB} \ln\left(\frac{M_\mathrm{GUT}}{M_K}\right)~,
\label{RGEsol1}
\end{equation}
can lead to lepton-flavour violating (LFV) radiative decays 
of charged leptons, e.g.~$\mu\to e\gamma$, $\tau\to e\gamma$ and
$\tau\to\mu\gamma$\cite{borzumati86}. 
Assuming, for simplicity, that there
is a common mass scale $M$ of the right-chiral neutrinos and that
there are no large cancellations in the seesaw formula, 
Eq.~\ref{seesawfor}, one can estimate
the LFV rates. Neglecting $\mathcal{O}(1)$ factors,
as well as the logarithmic factor in Eq.~\ref{RGEsol1}, one obtains:
\beq
\frac{\mathrm{BR}(\ell_A\to \ell_B\gamma)}{\mathrm{BR}(\ell_A\to \ell_B\bar{\nu}_B\nu_A)} \approx 10^{-7} \times \left( \frac{100\gev}{m_\mathrm{SUSY}}\right)^4 \left( \frac{\tan\beta}{10}\right)^2 \left( \frac{m_{\nu_3}}{5\times 10^{-2}\,\mathrm{eV}}\right)^2 \left( \frac{M}{10^{14}\gev}\right)^2~.
\label{brest}
\eeq
No such decays have been observed so far,
but the upcoming experiments aim at searching for
these processes with high accuracy. The current experimental bounds
for the LFV radiative decays and related $\mu\to e$ conversion 
in nuclei, as well as the sensitivities of planned experiments
are presented in Table \ref{table.lfv.exp}.

\begin{table}[t]
\caption{Current and expected sensitivities (90\% CF) of experiments searching for LFV.
\label{table.lfv.exp}}
\begin{center}
\begin{tabular}{|c|cc|ccc|}
\hline
process & \multicolumn{5}{|c|}{experimental constraints} \\
& \multicolumn{2}{|c|}{current} & \multicolumn{3}{|c|}{planned} \\
%& & & & & \\
& result & experiment & sensitivity & year & experiment \\
\hline
$\mu^+\to e^+\gamma$ & $1.2\times 10^{-11}$ & {\em MEGA-LAMPF} \cite{brooks99} & $10^{-13}$ & 2006 & {\em MEG} \cite{aoki04}\\
$\mu^-\mathrm{Ti}\to e^-\mathrm{Ti}$ & $6.1\times 10^{-13}$ & {\em SINDRUM II} \cite{wintz98} & & & \\
$\mu^-\mathrm{Pb}\to e^-\mathrm{Pb}$ & $4.6\times 10^{-11}$ & {\em SINDRUM II} \cite{honecker96} & & & \\
$\mu^-\mathrm{Au}\to e^-\mathrm{Au}$ & $8\times 10^{-13}$ & {\em SINDRUM II} \cite{bertl02} & & & \\
$\mu^-\mathrm{Al}\to e^-\mathrm{Al}$ & & & $2\times 10^{-17}$ & 2009 & {\em MECO} \cite{aoki04}\\
\hline
$\teg$ & $3.7\times 10^{-7}$ & {\em Belle} \cite{inami03} & $10^{-9}$ & 2007 & {\em LHC} \cite{unel05}\\
\hline
$\tmg$ & $6.8\times 10^{-8}$ & {\em BABAR} \cite{wilson05} & $10^{-9}$ & 2007 & {\em LHC} \cite{unel05}\\
\hline
\end{tabular}
\end{center}
\end{table}

The specific predictions for the LFV decays 
(as well as for leptogenesis)
depend crucially on the parameters of the seesaw model which are
not constrained by neutrino experiments\cite{davidson03}.
There is a very wide range of predictions for the rates
of the LFV decays, which, as follows from Eq.~\ref{brest}, 
depend on the masses of the right-chiral neutrinos,
on the textures of the neutrino Yukawa couplings and on the parameters
of the MSSM. 
In principle, more specific predictions can be obtained
under an additional assumption of successful leptogenesis, possibly
consistent with the gravitino bounds.

\section{Leptogenesis and LFV decays}

If the right-chiral neutrinos have hierarchical masses, then
the lepton asymmetry produced in leptogenesis is directly
proportional to the $CP$ asymmetry $\varepsilon_1$ 
in the decays of the lightest right-chiral neutrino.
Under reasonable assumptions about the neutrino Yukawa couplings,
this $CP$ asymmetry must satisfy the Davidson-Ibarra 
bound\cite{davidson02}:
\beq
|\varepsilon_1| \leq \varepsilon_1^\mathrm{DI} = \frac{3}{8\pi} \frac{M_1(m_{\nu_3}-m_{\nu_1})}{\langle H\rangle^2}
\label{DIbound}
\eeq
Successful leptogenesis requires
sufficiently large $\varepsilon_1$, which, 
as follows from Eq.~\ref{DIbound}, is possible if
the mass 
$M_1$ of the lightest right-chiral neutrino is sufficiently large.
Interestingly, the case of maximally efficient leptogenesis
corresponds to the decoupling of the lightest right-chiral
neutrino and
$T_\mathrm{RH}\simgt M_1\simgt 10^9\gev$.
As regards models with one of the heavier 
right-chiral neutrino decoupled,
successful leptogenesis generically requires
$5T_\mathrm{RH}>M_1\simgt 10^{11}\gev$,
due to stonger washout of the generated lepton asymmetry
and a slight suppression of the maximal $CP$ 
asymmetry\cite{chankowski03}.
The first case is marginally consistent
with the weak gravitino bound, but there are no additional
constraints on the predictions
for LFV decays. In the second case, the requirement of successful
thermal leptogenesis 
implies that $M_1$ is so large that $\mu\to e\gamma$ should be, 
in principle, observable in the upcoming 
experiments\footnote{Throughout this work, 
we do not mention the predictions for
the rates of LFV $\tau$ decays, since they are similar to those of 
$\mu\to e\gamma$, but the relevant experiments are less sensitive.
}; this case is, however,
inconsistent with both gravitino bounds.
% end of new part 

\begin{figure}[t]
\caption{Extremal values of $\bmeg/f(m_0,M_{1/2})$ as a 
function of $|z|$ for $\sin\theta_{13}=0$ (panel a) and 
$\sin\theta=0.1,\delta=-\pi/2$ (panel b), $\tan\beta =10$ 
and $A_0=0$. 
We have chosen $f(m_0,M_{1/2})=1$ for values $m_0=100\gev$, 
$M_{1/2}=500\gev$, consistent with the dark matter relic 
density.
%\cite{ellis03}.
Dotted, dashed and solid curves 
correspond to $\mathrm{arg}\, z=0$, $\pi/4$ and $\pi/2$, respectively.
The masses of the right-chiral neutrinos are 
$M_{1,2,3}=(2,10,50)\times 10^{13}\gev$.
The values of the remaining mixing angles are set to their
central values and the neutrino Majorana phases vary freely.
The horizontal line represents the current experimental
bound. \label{fig.lfv.pred}}
\includegraphics*[height=7cm]{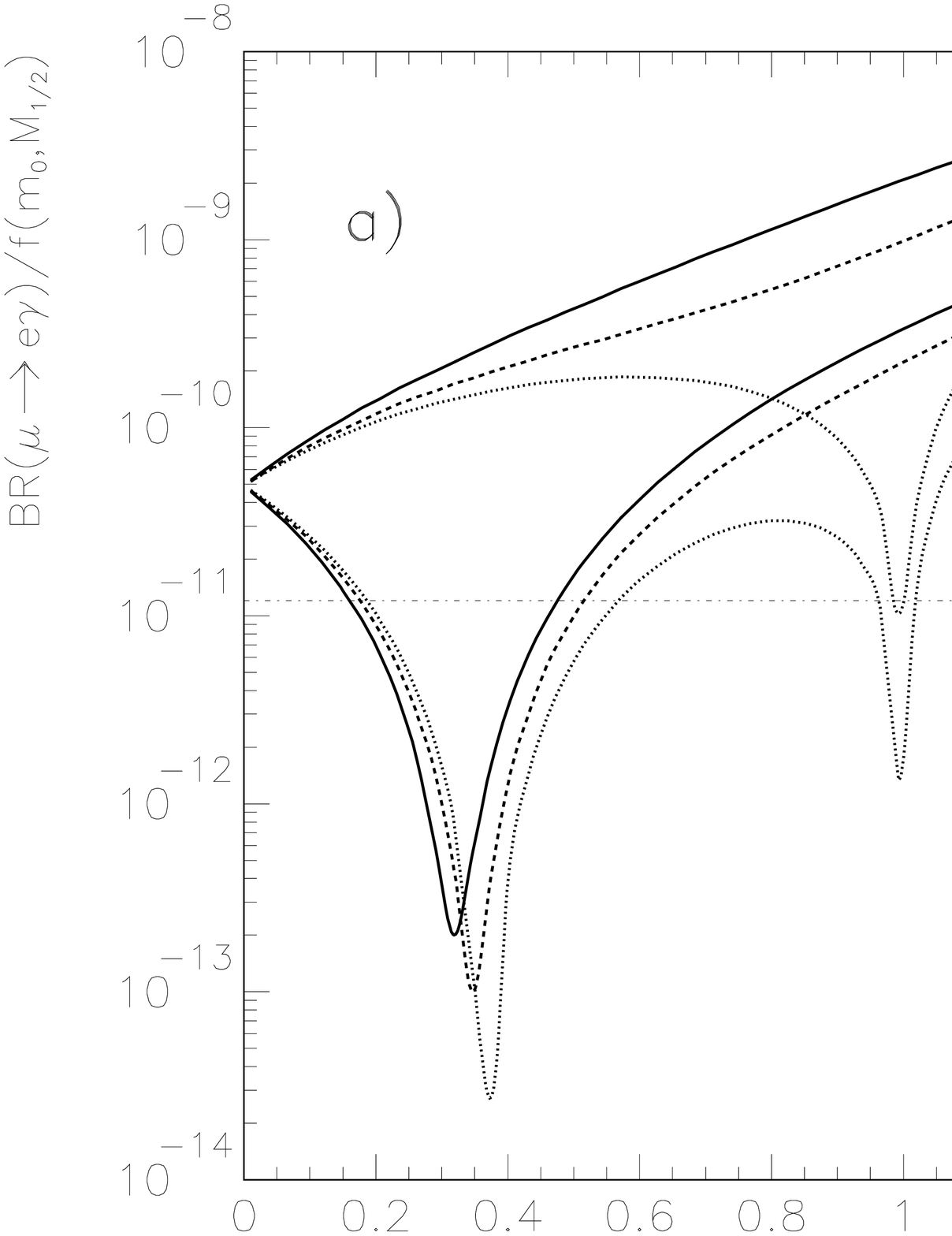}
\hspace{2cm}
\includegraphics*[height=7cm]{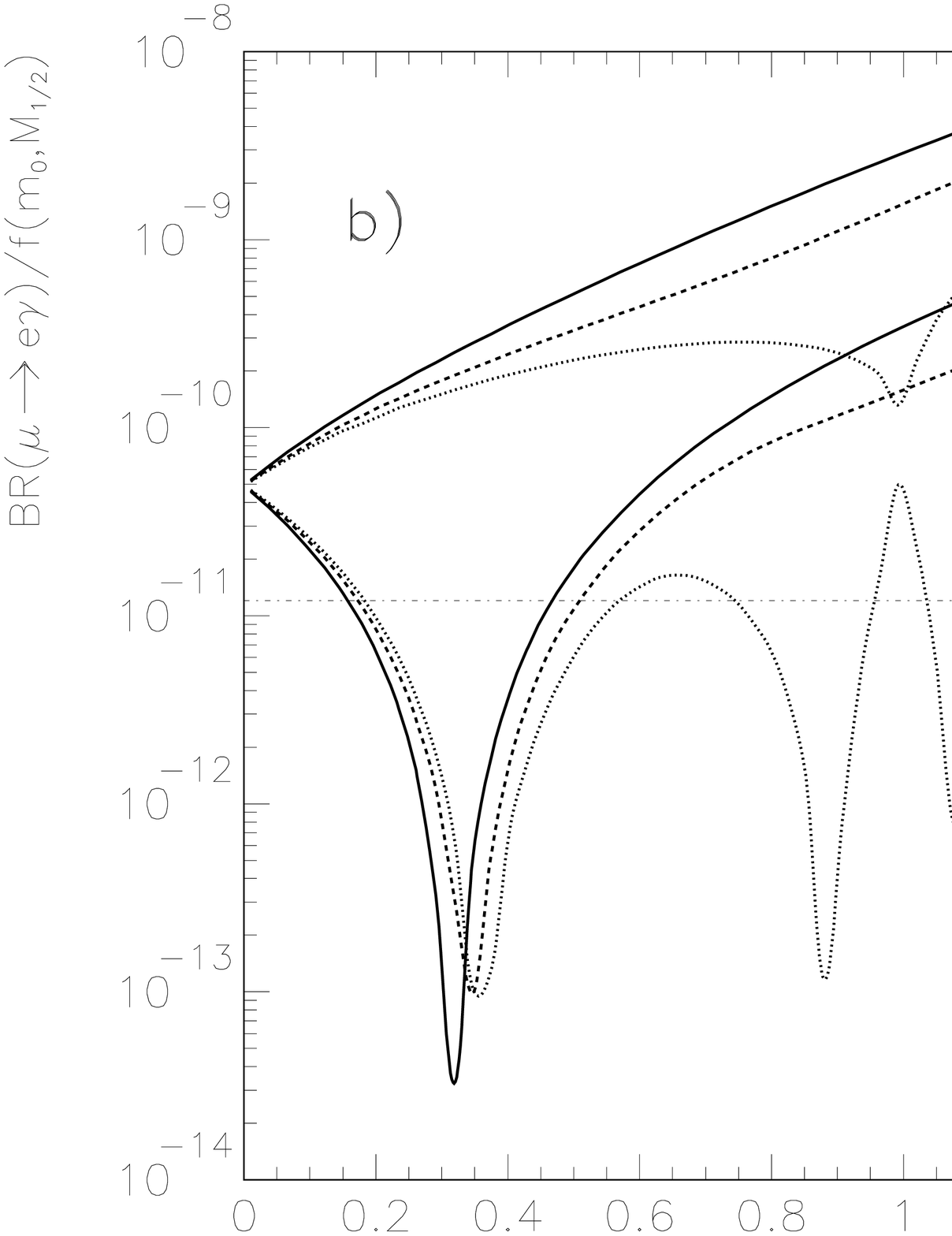}
\end{figure}

Various alterations of generic neutrino mass
models have been proposed to allow for successful leptogenesis
with a low reheating temperature.
In particular, for the lightest right-chiral
neutrino decoupled, it has been shown that the 
simplest version of nonthermal
leptogenesis embedded in the scenario of 
sneutrino-driven inflation\cite{ellis04}
can be successful for $T_\mathrm{RH}\sim10^6\gev$. The right-chiral 
(s)neutrinos are rather heavy in this scenario, as the temperature
fluctuations of the cosmic microwave background require 
$M_1=2\times 10^{13}\gev$, and, according to Eq.~\ref{brest}, 
the model 
predicts a large rate of $\mu\to e\gamma$\cite{chankowski04}, 
possibly observable in the
upcoming experiments, as shown in Figure \ref{fig.lfv.pred}.
One can also construct a model,
in which the $CP$ asymmetry in the decays of the lightest right-chiral
neutrino is enhanced due to large Yukawa couplings of the heavy 
right-chiral neutrinos\cite{raidal05}.
In this model, successful thermal leptogenesis can be achieved for
arbitrarily low $T_\mathrm{RH}$ and the model actually implies that
$\mu\to e\gamma$ will be observed in the forthcoming experiments,
unless the supersymmetric particles are very heavy.
For the heaviest right-chiral neutrino decoupled, 
successful leptogenesis is possible for $T_\mathrm{RH}$ consistent
with the strong gravitino bound, if the non-decoupled right-chiral
neutrinos are tightly degenerate in mass\cite{flanz95,chankowski03}. 
In particular, it has been proposed that this quasi-degeneracy
results, via renormalization group corrections, 
from an exactly degenerate mass spectrum of the right-chiral
neutrinos at the scale of Grand Unification\cite{gonzalez04}.
Such model is, however, subject to certain 
consistency conditions\cite{turzynski04}, 
partially
compensating this enhancement of the $CP$ asymmetry,
but successful leptogenesis is, nevertheless, possible for $T_\mathrm{RH}$
consistent with the strong gravitino bound, if $\tan\beta\simgt10$.

\section{Summary}

Let us conclude this presentation by an observation that,
if sparticle mass spectrum consistent with the hypothesis of universal
soft masses and dark matter constraints is measured at the {\em LHC}, and
if $\mu\to e\gamma$ is observed, this would be a strong argument that
the seesaw mechanism is correct and the strong gravitino bound on the
reheating temperature holds. 
Analogous conclusions can also be drawn from a confirmed observation
of neutrinoless double beta decay. 
Then, if the baryon asymmetry of the Universe
is to be generated in leptogenesis, it will be interesting to look
for a symmetry principle, which could justify special features
of the seesaw neutrino mass models with a right-chiral 
neutrino decoupled, presented in this work.

\section*{Acknowledgments}
K.~T.~thanks S.~Pokorski for continuous support and encouragement, 
and P.\ H.\ Chankowski for many useful discussions.
K.~T.~is partially supported by the Polish State 
Committee for Scientific Research Grant 
1~P03D~014~26 for years 2004-06.

\end{document}